\documentclass[preprint,prd,amsmath,amssymb,aps,nofootinbib]{revtex4-1}

\usepackage{graphicx}
\usepackage{siunitx}
\usepackage{bm}
\usepackage{natbib}
\def\mathbi#1{\textbf{\em #1}}

\def\apj{Astrophys.\ J.\ }

\def\ssr{Space\ Sci.\ Rev.\ }

\begin{document}
\title{Features of the gamma-ray pulsar halo HESS~J1831$-$098}
\author{Kun Fang$^{a}$}
\author{Shao-Qiang Xi$^{a}$}
\author{Li-Zhuo Bao$^{a,b}$}
\author{Xiao-Jun Bi$^{a,b}$}
\author{En-Sheng Chen$^{a,b}$}

\affiliation{
$^a$Key Laboratory of Particle Astrophysics, Institute of High Energy
Physics, Chinese Academy of Sciences, Beijing 100049, China \\
$^b$University of Chinese Academy of Sciences, Beijing 100049, China\\
}

\date{\today}

\begin{abstract}
Gamma-ray pulsar halos are ideal indicators of cosmic-ray propagation in localized regions of the Galaxy and electron injection from pulsar wind nebulae. HESS~J1831$-$098 is a candidate pulsar halo observed by both H.E.S.S. and HAWC experiments. We adopt the flux map of the H.E.S.S. Galactic plane survey and the spectrum measurements of H.E.S.S. and \textit{Fermi}-LAT to study HESS~J1831$-$098. We find that HESS~J1831$-$098 meets all the criteria for a pulsar halo. The diffusion coefficient inside the halo and the conversion efficiency from the pulsar spin-down energy to the electron energy are both similar to the Geminga halo, a canonical pulsar halo. The injection spectrum can be well described by an exponentially-cutoff power law. However, the needed power-law term is very hard with $p\lesssim1$ if the diffusion coefficient is spatially and temporally independent. Considering the possible origins of the slow-diffusion environment, we adopt the two-zone diffusion model and the time-delayed slow-diffusion model. Both the models can interpret the H.E.S.S. and \textit{Fermi}-LAT results with a milder $p$. A modified injection time profile may have a similar effect.
\end{abstract}

\maketitle

\section{Introduction}
\label{sec:intro}
Pulsar halos are inverse Compton (IC) gamma-ray sources generated by electrons and positrons\footnote{\textit{Electrons} will denote both electrons and positrons hereafter if not specified.} escaping from pulsars/pulsar wind nebulae (PWNe) and diffusing very inefficiently in the interstellar medium (ISM) around pulsars \cite{Linden:2017vvb,Sudoh:2019lav,Giacinti:2019nbu}. They are ideal indicators of cosmic-ray (CR) propagation in localized regions of the Galaxy as the gamma-ray morphologies of the halos unambiguously trace the spatial distributions of the parent electrons. Observed pulsar halos show that the diffusion coefficient near the sources is more than two orders of magnitude smaller than the typical value in the Galaxy \cite{Abeysekara:2017old,Aharonian:2021jtz}, which has a significant impact on the study of Galactic CR propagation. Meanwhile, spectral measurements of pulsar halos may provide a unique estimation for the electron injection spectrum of PWNe \cite{Fang:2022mdg}.

Pulsar halos observed at present include the Geminga halo, the Monogem halo, and LHAASO~J0622$+$3755 \cite{Abeysekara:2017old,Aharonian:2021jtz}. The third catalog of the High-Altitude Water Cherenkov (HAWC) observatory also lists six other candidate pulsar halos \cite{Albert:2020fua}. However, most of the current spectral measurements for pulsar halos are above $\sim10$~TeV, where the spectrum is believed to be dominated by the high-energy cutoff term, possibly due to the PWN acceleration limit. We can hardly understand the features of the injection spectrum below the high-energy cutoff with the spectrum above $\sim10$~TeV alone. On the other hand, the low-energy spectrum of pulsar halos may provide richer information for the CR propagation in the halo region \cite{Fang:2021qon}.

The energy spectrum measurement of the High Energy Spectroscopic System (H.E.S.S.) covers $0.1\lesssim E_\gamma\lesssim100$~TeV, which may provide us with a clearer understanding of the injection spectral features of pulsar halos. The H.E.S.S. preliminary measurement of the Geminga halo indicates an unexpected low-energy spectral component \cite{Mitchell:2021tig}, possibly generated by a new electron population \cite{Fang:2022mdg}. H.E.S.S. also has advantages in studying distant pulsar halos owing to the high angular resolution. There are two intersections between the H.E.S.S. Galactic plane survey (HGPS, \cite{HESS:2018pbp}) and the pulsar halo catalog of HAWC: HESS~J1831$-$098 (3HWC~J1831$-$095) and HESS~1912$+$101 (3HWC~1912$+$103). The origin of HESS~1912$+$101 is still unclear \cite{Aharonian:2008uj,2020ApJ...889...12Z,Sun:2022tcn}, while HESS~J1831$-$098 is very likely a pulsar halo associated with the powerful middle-aged pulsar PSR~J1831$-$0952, as we will show below. Apart from the morphology information given by the HGPS sky map, a spectral measurement of HESS~J1831$-$098 in $0.3-30$~TeV is also available \cite{Sheidaei:2011vg}.

In this work, we study the features of the pulsar halo HESS~J1831$-$098 with the H.E.S.S and \textit{Fermi} Large Area Telescope (\textit{Fermi}-LAT) observations. \textit{Fermi}-LAT provides spectral measurements below the H.E.S.S. energy range, which can give further constraints to models. Sec.~\ref{sec:model} introduces the basic model of pulsar halos, including the characteristics of pulsar halos and the calculation of the gamma-ray emission. In Sec.~\ref{sec:fermi}, we present the data analysis of \textit{Fermi}-LAT. In Sec.~\ref{sec:hess}, we constrain the model parameters by fitting the gamma-ray morphology and spectrum of HESS~J1831$-$098 measured by H.E.S.S. and compare the result with the flux upper limits (ULs) of \textit{Fermi}-LAT. Under the basic model, the H.E.S.S. and \textit{Fermi}-LAT data may not be interpreted with a reasonable injection spectral index. We discuss more sophisticated models in Sec.~\ref{sec:discuss} to reconcile the conflicts. Sec.~\ref{sec:conclu} is the conclusion and prospect. 

\section{Basic model of pulsar halos}
\label{sec:model}
After the accelerated electrons escape from the PWNe of a middle-aged pulsar\footnote{Millisecond pulsars may also have pulsar halos as pointed out by Ref.~\cite{Hooper:2018fih,Hooper:2021kyp}.}, they freely propagate in the surrounding ISM. The propagation of the electrons is generally considered the diffusion process. If the diffusion coefficient is small, the electrons will accumulate around the pulsar and generate an observable gamma-ray halo, namely the pulsar halo, through the IC scattering of the background photons. The origin of the slow-diffusion environment is still unclear. It may be self-excited by the escaping electrons \cite{Evoli:2018aza,Mukhopadhyay:2021dyh}, while the electron energy may not be enough to suppress the diffusion coefficient to the required level \cite{Kun:2019sks}. It may also be a pre-existing turbulent region left by the parent supernova remnant (SNR) of the pulsar or other anterior energy injection processes \cite{Kun:2019sks}. Anisotropic diffusion is also proposed to account for the Geminga halo \cite{Liu:2019zyj}, while it may not be a general interpretation of this class of sources \cite{DeLaTorreLuque:2022chz}.

According to the characteristics of pulsar halos, we present our criteria (CT) for a pulsar halo as follows.
\begin{enumerate}
    \item A pulsar halo should have spatial coincidence with a powerful pulsar. More specifically, the pulsar should be located around the centroid of the pulsar halo, at least for $E_\gamma\gtrsim1$~TeV. The pulsar may be off-center at lower energies owing to its proper motion \cite{DiMauro:2019hwn,ZhangZhangYi:2021kzq}.
    \item The spin-down luminosity of the pulsar should be large enough to generate the pulsar halo. This criterion is not only essential for a pulsar halo, but also for the pulsar halo models \cite{2021arXiv210707395B}.
    \item We set a lower limit of 50~kyr for the pulsar age, which may exclude a pure PWN or a source in the mixed state of a pulsar halo and a relic PWN. The 3HWC catalog adopts a more conservative lower limit of 100~kyr \cite{Albert:2020fua}.  
    \item The extension of a gamma-ray pulsar halo should be significantly larger than that of the x-ray PWN (if the x-ray PWN is observable), which is also essential to distinguish between a pulsar halo and a PWN. The PWN of a middle-aged pulsar should be in the bow-shock phase and has a small extension of $\lesssim1$~pc \cite{Gaensler:2006ua}, compared with the $\gtrsim10$~pc extension for a pulsar halo.
\end{enumerate}

We will show in Sec.~\ref{sec:hess} that HESS~J1831$-$098 meets CT~1 and 2. The age of PSR J1831$-$0952 is $t_p=128$~kyr as given by the Australia Telescope National Facility (ATNF) catalog \cite{Manchester:2004bp}, satisfying CT~3. A x-ray source at the position of PSR J1831$-$0952 is detected by the \textit{Chandra} x-ray observatory, which is likely the PWN of PSR J1831$-$0952 \cite{2019ATel12463....1A}. The extension of the possible PWN is reported to be several arcsec, corresponding to $\sim0.1$~pc given the pulsar distance $d_p=3.68$~kpc. Thus, CT~4 may also be satisfied. Kes~69 is a cloud-interacting SNR located south of HESS~J1831$-$098 \cite{Zhou:2008np} and is the closest SNR to HESS~J1831$-$098 on the sky map. However, the lack of gamma-ray emission from Kes~69 implies that it may not be a powerful accelerator of high-energy CRs \cite{2011PhDT.......329B,Sezer:2018qgb}, and HESS~J1831$-$098 is unlikely to be lit up by CR nuclei escaping from Kes~69. All these indicate that HESS~J1831$-$098 could be reasonably a pulsar halo.

Then we briefly introduce the calculation of the gamma-ray emission of HESS~J1831$-$098. The electron propagation equation can be expressed by
\begin{equation}
  \frac{\partial N(E_e, \mathbi{r}, t)}{\partial t} = \nabla \cdot[D(E_e)\nabla N(E_e, \mathbi{r}, t)] + \frac{\partial[b(E_e)N(E_e, \mathbi{r}, t)]}{\partial E_e} + Q(E_e, \mathbi{r}, t)\,,
 \label{eq:prop}
\end{equation}
where $N$ is the electron number density, and $E_e$ is the electron energy. The diffusion coefficient takes the form of $D(E_e)=D_{100}(E_e/{\rm 100~TeV})^\delta$ and is assumed to be spatially and temporally independent in the basic model. We set the diffusion coefficient at 100~TeV as the pivot in order to compare it with the result of other pulsar halos. We set $\delta=1/3$ as suggested by Kolmogorov's theory. The second and third terms on the right-hand side are the energy-loss and source terms, respectively, where $b(E_e)$ is the electron energy-loss rate.

Synchrotron radiation and IC scattering dominate the energy losses of high-energy electrons. We take the magnetic field strength of $B=3$~$\mu$G for the synchrotron loss rate. The seed photon field of IC scattering consists of the cosmic microwave background (CMB), the infrared dust emission, and the starlight. The temperature and energy density of CMB are 2.725~K and 0.26~eV~cm$^{-3}$ \cite{Fixsen:2009ug}, respectively. We adopt the methods introduced in Ref.~\cite{Vernetto:2016alq} to get the infrared and starlight components. 
We simplify the infrared and starlight components by searching for their best-fit gray-body distributions. Considering the position of PSR J1831$-$0952, the temperatures and energy densities of the infrared and starlight components are 31~K, 0.48~eV~cm$^{-3}$ and 4300~K, 1.3~eV~cm$^{-3}$, respectively. The parameterization method given in Ref.~\cite{Fang:2020dmi} is used to calculate the IC energy-loss rate. 

The source function takes the form of
\begin{equation}
 Q(E_e,\mathbi{r},t)=\left\{
 \begin{aligned}
 & q(E_e)\,\delta(\mathbi{r}-\mathbi{r}_p)\,[(t_p+t_ { \rm
sd})/(t+t_{\rm sd})]^2\,, & t\geq0 \\
 & 0\,, & t<0
 \end{aligned}
 \right.\,,
 \label{eq:src}
\end{equation}
where $q(E_e)$ is the current electron injection spectrum, $\mathbi{r}_p$ and $t_p$ are the position and age of PSR~J1831$-$0952, respectively, and $t_{\rm sd}$ is the pulsar spin-down time scale, which is set to be 10~kyr. The time profile of the source function is assumed to follow the pulsar spin-down luminosity, and $t=0$ corresponds to the birth time of the pulsar.

The injection spectrum is assumed to be a cutoff power law as
\begin{equation}
 q(E_e)\propto E_e^{-p}\,{\rm exp}\left[-\left(\frac{E_e}{E_c}\right)^2\right]\,.
 \label{eq:inj}
\end{equation}
The cutoff term describes the acceleration limit of the PWN, the form of which is suggested by Ref.~\cite{Zirakashvili:2006pv}. The normalization of the injection spectrum can be obtained by the relation of $\int_{\rm 1GeV}^{\infty} q(E_e)E_edE_e=\eta L$, where $L=1.08\times10^{36}$~erg~s$^{-1}$ is the current pulsar spin-down luminosity \cite{Manchester:2004bp}, and $\eta$ is the conversion efficiency from the spin-down energy to the electron energy. 

Eq.~(\ref{eq:prop}) can be solved with the Green's function method. We integrate $N$ over the line of sight from Earth to the vicinity of the pulsar and then obtain the gamma-ray surface brightness of the halo $S(\theta,E_\gamma)$ with the standard calculation of IC scattering \cite{Blumenthal:1970gc}, where $\theta$ is the angle away from the pulsar. One may refer to the previous works (e.g., Ref.~\cite{Fang:2022mdg}) for details of the above calculation.

\section{Analysis of \textit{Fermi}-LAT data}
\label{sec:fermi}
The \textit{Fermi}-LAT is a pair conversion telescope covering the energy range between 20~MeV and $>500$~GeV \cite{Fermi-LAT:2009ihh}.  In this study, we undertake the \textit{Fermi}-LAT analysis employing the  \textit{Fermipy} python package which is built on the standard LAT analysis software Fermi Science Tools. We use the data collected over a period of approximately 13 year with the Pass 8 response functions expressed by P8R3\_SOURCE\_V3. In order to take advantage of the improved LAT resolution and reduced the contamination of the pulsar, we use the event above 10~GeV. We reconstruct a $\ang{16} \times \ang{16}$ region of interesting (ROI) around the position of PSR~J1831-0952, select time intervals with good data quality and exclude the events with zenith angle larger than $\ang{90}$ to limit contamination from the gamma-ray-bright Earth limb. We use the background model including the Galactic diffuse emission (gll\_iem\_v07.fits) and isotropic emission (iso\_P8R3\_SOURCE\_v3\_v1.txt), as well as the point-like and extended sources in the fourth \textit{Fermi}-LAT catalog \cite{Fermi-LAT:2019yla}. We optimize the spectral parameters of each sources in the background model using the \textit{optimize} tool in the \textit{Fermipy} package. We search for the possible gamma-ray excess associated with HESS~J1831-098 using the test statistic(TS) map with a test source modeled by a $\ang{0.3}$ disk spatial template. No detection is found, and the 95\% ULs are derived using the $\ang{0.3}$ disk model centered at (${\rm l}=\ang{21.86}$, ${\rm b}=\ang{-0.051}$), which are shown in Fig.~\ref{fig:spec}.

\section{Constraining parameters with H.E.S.S. and \textit{Fermi}-LAT data}
\label{sec:hess}
The diffusion coefficient and injection spectral parameters are the main parameters for the study of a pulsar halo. The HGPS flux map can be used to estimate the diffusion coefficient of HESS~J1831$-$098, while the gamma-ray spectral measurement of HESS~J1831$-$098 can effectively constrain its electron injection spectrum.  

\begin{figure}[t]
\centering
\includegraphics[width=0.68\textwidth]{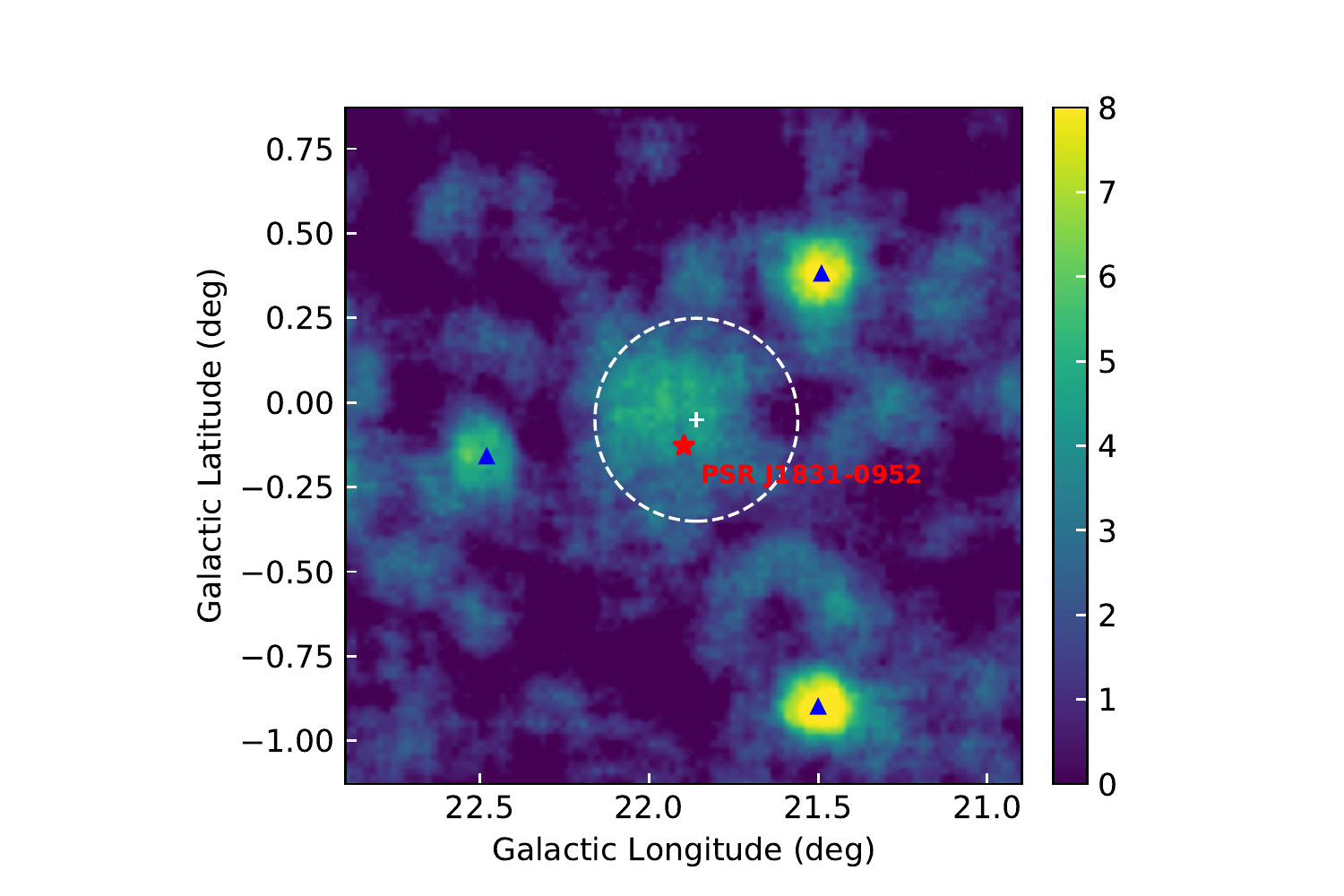}
\caption{HGPS significance map ($R_c=\ang{0.1}$) for the region around HESS~J1831$-$098. The white cross marks the fitted position of HESS~J1831$-$098 assuming a Gaussian template, while the white dotted circle shows the fitted extent of $\sigma_{\rm src}=\ang{0.30}$. The red star shows the position of the associated pulsar PSR~J1831$-$0952. The blue triangles show the positions of the surrounding bright sources given by the H.E.S.S. catalog \cite{HESS:2018pbp}.}
\label{fig:signi}
\end{figure}

Fig.~\ref{fig:signi} shows the HGPS significance map for the region around HESS~J1831$-$098, with an integration radius of $R_c=\ang{0.1}$. This $\ang{2}\times\ang{2}$ region centered at the position of PSR~J1831$-$0952 is considered the ROI for the morphology study. It can be seen that HESS~J1831$-$098 is surrounded by three bright sources without significant extension, i.e., HESS~J1828$-$099, HESS~J1832$-$093, and HESS~J1833$-$105, which may affect the estimation of the diffusion coefficient for HESS~J1831$-$098. To estimate the contribution from these three sources, we use a template of four 2D Gaussian functions to fit the flux map ($R_c=\ang{0.1}$) within the ROI as the preparation step. The morphologies of the three point-like sources are assumed to follow the point spread function (PSF) of H.E.S.S., while the specific form of the PSF is not available. We approximate the PSF by a single 2D Gaussian function, the extension of which, $\sigma_{\rm PSF}$, is determined by the data fit. The positions of the three surrounding sources are taken from the HGPS source catalog. For HESS~J1831$-$098, the intrinsic extension $\sigma_{\rm src}$ and central coordinates, (l$_{\rm src}$, b$_{\rm src}$), are set as free parameters. Thus, the preparation step has eight free parameters, including the flux normalizations of the four sources. We compute the spatially correlated template with $R_c=\ang{0.1}$ to compare it with the H.E.S.S. data.

The best-fit parameters given by the maximum likelihood method are: $\sigma_{\rm PSF}=\ang{0.076}$, $\sigma_{\rm src}=\ang{0.30}$, ${\rm l}_{\rm src}=\ang{21.86}$, and ${\rm b}_{\rm src}=\ang{-0.051}$. The 68\% containment radius of the PSF model given by the HGPS paper varies from $\sim\ang{0.08}$ to $\sim\ang{0.15}$ depending on different field-of-view offsets \cite{HESS:2018pbp}. It corresponds to $\sigma\sim\ang{0.05}-\ang{0.10}$ for a 2D Gaussian PSF, which is consistent with our best-fit $\sigma_{\rm PSF}$. We adopt this best-fit $\sigma_{\rm PSF}$ in the following calculations. The best-fit $\sigma_{\rm src}$ can be a straightforward estimate for the source extension. It is significantly broader than the PSF, consistent with the nature of HESS~J1831$-$098 as a pulsar halo. The angular distance between the best-fit centroid of HESS~J1831$-$098 and the coordinates of PSR~J1831$-$0952, ($\ang{21.897}$, $\ang{-0.128}$), is $\ang{0.08}$. It means that HESS~J1831$-$098 is in good agreement with the pulsar in position, meeting CT~1 in Sec.~\ref{sec:model}. 

We fix the fluxes of the three surrounding sources obtained in the preparation step and adopt the basic model introduced in Sec.~\ref{sec:model} for HESS~J1831$-$098 to fit the flux map. The HGPS flux map provides the integrated fluxes above 1~TeV, and we compute $\int_{\rm 1TeV}^{\infty}S(\theta,E_\gamma)E_\gamma dE_\gamma$ to compare with the data. Note that $S(\theta,E_\gamma)$ needs to be convoluted with the PSF. The free parameters are: l$_{\rm src}$, b$_{\rm src}$, $D_{100}$, and a normalization parameter. The injection spectrum also has a little effect on the morphology. Meanwhile, the gamma-ray spectrum of HESS~J1831$-$098 measured by Ref.~\cite{Sheidaei:2011vg} is within a circular region of $\ang{0.3}$ from the centroid, the theoretical value of which depends on both the injection spectrum and the diffusion coefficient. Thus, we iteratively fit the flux map and the gamma-ray spectrum until stable results are obtained. The main fitting results are summarized in Table~\ref{tab:para}.

\begin{table}[t]
 \centering
 \caption{Constraints on the main parameters.}
 \begin{tabular}{lcc}
  \hline
  \hline
  Parameter & Best-fit value & \quad 68\% posterior CI \quad \\
  \hline
  $p$ & 0.88 & [-1.05, 1.08] \\
  \hline
  $E_c$ (TeV) & 52 & [26, 58] \\
  \hline
  $\eta$ (\%) & 6.6 & [5.2, 7.2] \\
  \hline
  \hline
  $D_{100}$ (cm$^2$~s$^{-1}$) & \quad $9.0\times10^{27}$ \quad & - \\
  \hline
 \end{tabular}
 \label{tab:para}
\end{table}

The best-fit parameters of the morphology fit are ${\rm l}_{\rm src}=\ang{21.90}$, ${\rm b}_{\rm src}=\ang{-0.046}$, and $D_{100}=9.0\times10^{27}$~cm$^2$~s$^{-1}$. The best-fit centroid is in good agreement with the pulsar position. The best-fit $D_{100}$ is consistent in magnitude with those of the Geminga halo, the Monogem halo, and LHAASO~J0621$+$3755, which are $3.2\times10^{27}$~cm$^2$~s$^{-1}$ \cite{Abeysekara:2017old}, $15\times10^{27}$~cm$^2$~s$^{-1}$ \cite{Abeysekara:2017old}, and $2.5\times10^{27}$~cm$^2$~s$^{-1}$ \cite{Fang:2021qon}, respectively. This strengthens the argument that HESS~J1831$-$098 is a pulsar halo. The measurement of the diffusion coefficient also depends on the pulsar distance. The distance of $d_p=3.68$~kpc is derived from the dispersion measure value of PSR~J1831$-$0952 and the electron density model given by Ref.~\cite{2017ApJ...835...29Y}. Ref.~\cite{Kutukcu:2022phy} argues that young pulsars ($t_p<800$~kyr) like PSR~J1831$-$0952 could still be located inside Galactic arms, and the distance of PSR~J1831$-$0952 is estimated to be $3.5\pm0.5$~kpc, consistent with the value we adopted. Thus, the conclusion should not be significantly affected by the uncertainty of the pulsar distance.

\begin{figure}[t]
\centering
\includegraphics[width=0.49\textwidth]{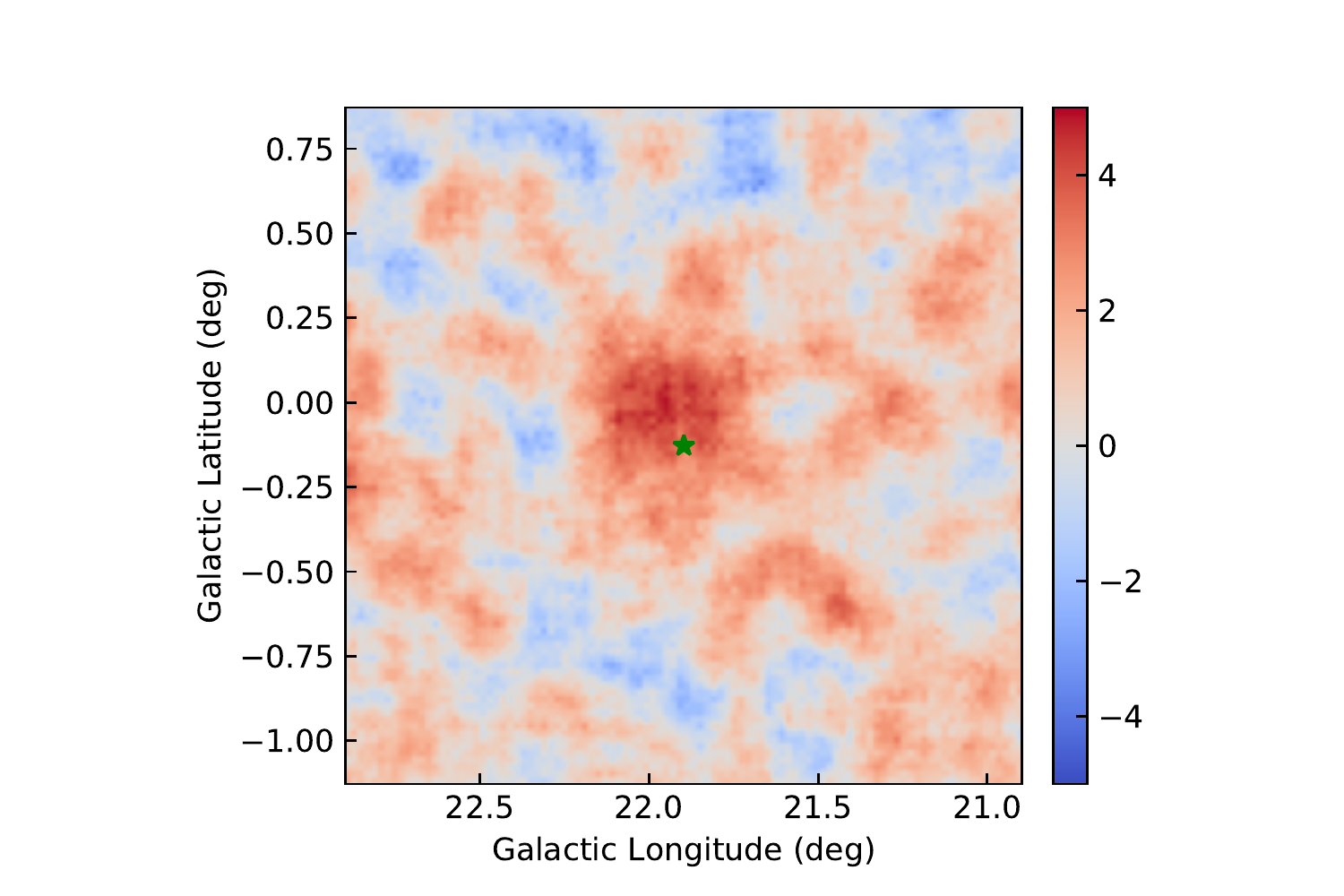}
\includegraphics[width=0.49\textwidth]{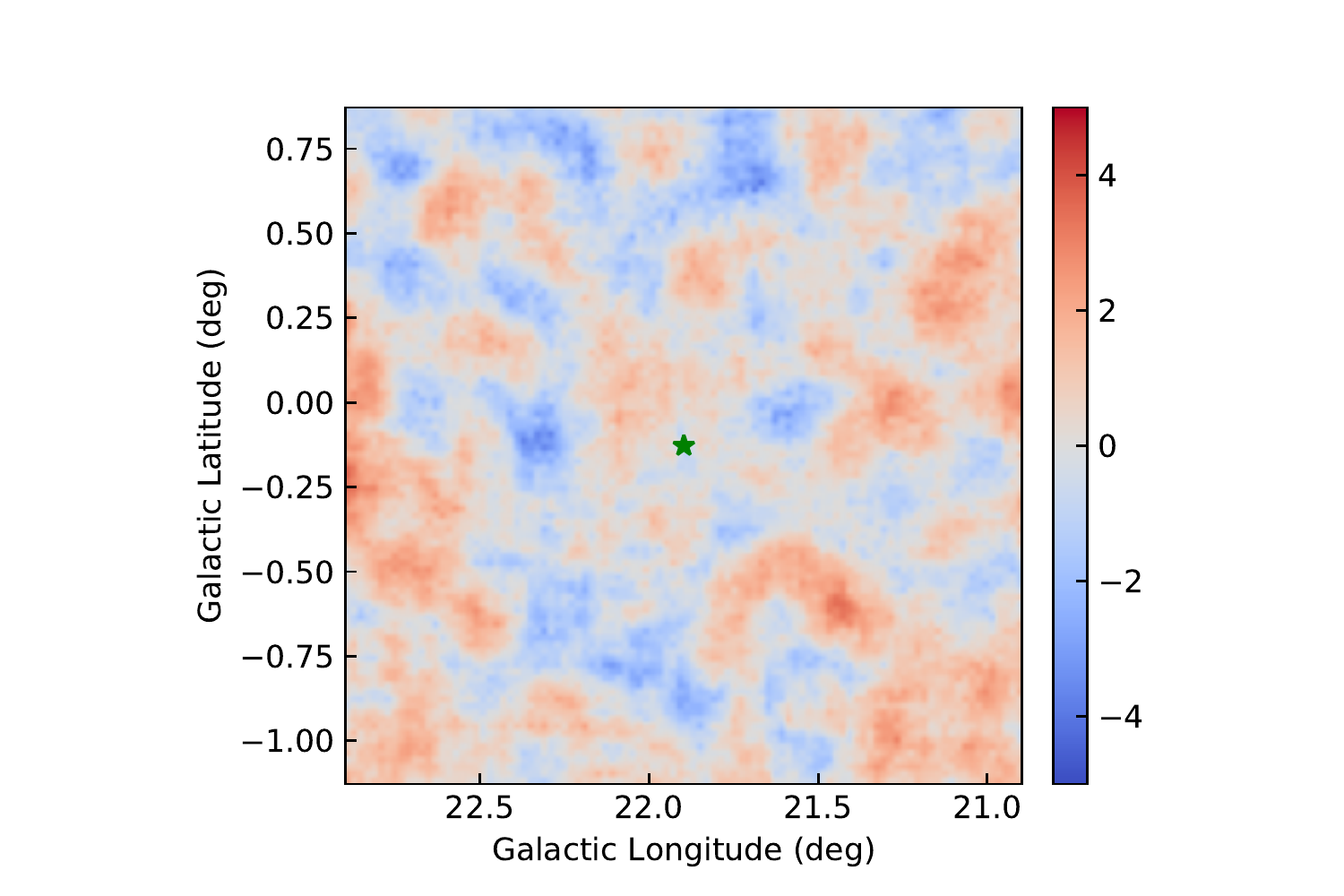}
\caption{Relative residual map $(F_{\rm exp}-F)/{\rm err}_{\rm exp}$ computing without (left) or with (right) the HESS~J1831$-$098 template, where $F_{\rm exp}$ and ${\rm err}_{\rm exp}$ are the flux and flux error with $R_c=\ang{0.1}$ given by the HGPS \cite{HESS:2018pbp}, respectively. The green star shows the position of PSR~J1831$-$0952.}
\label{fig:resid}
\end{figure}

In the morphology fits, we do not provide confidence intervals for the parameters. The released maps of HGPS are oversampled, which will lead to significant underestimates of the lengths of the confidence intervals. The goodness of fit cannot be provided for the same reason, while we give a comparison between the relative residual map $(F_{\rm exp}-F)/{\rm err}_{\rm exp}$ with and without the HESS~J1831$-$098 template in Fig.~\ref{fig:resid}, where $F_{\rm exp}$ and ${\rm err}_{\rm exp}$ are the flux and flux error released by H.E.S.S., respectively, and $F$ is the flux calculated with the model. It can be seen that the residual map in the ROI is quite homogeneous when the best-fit model of HESS~J1831$-$098 is considered, indicating that the source is well fitted.

\begin{figure}[t]
\centering
\includegraphics[width=0.6\textwidth]{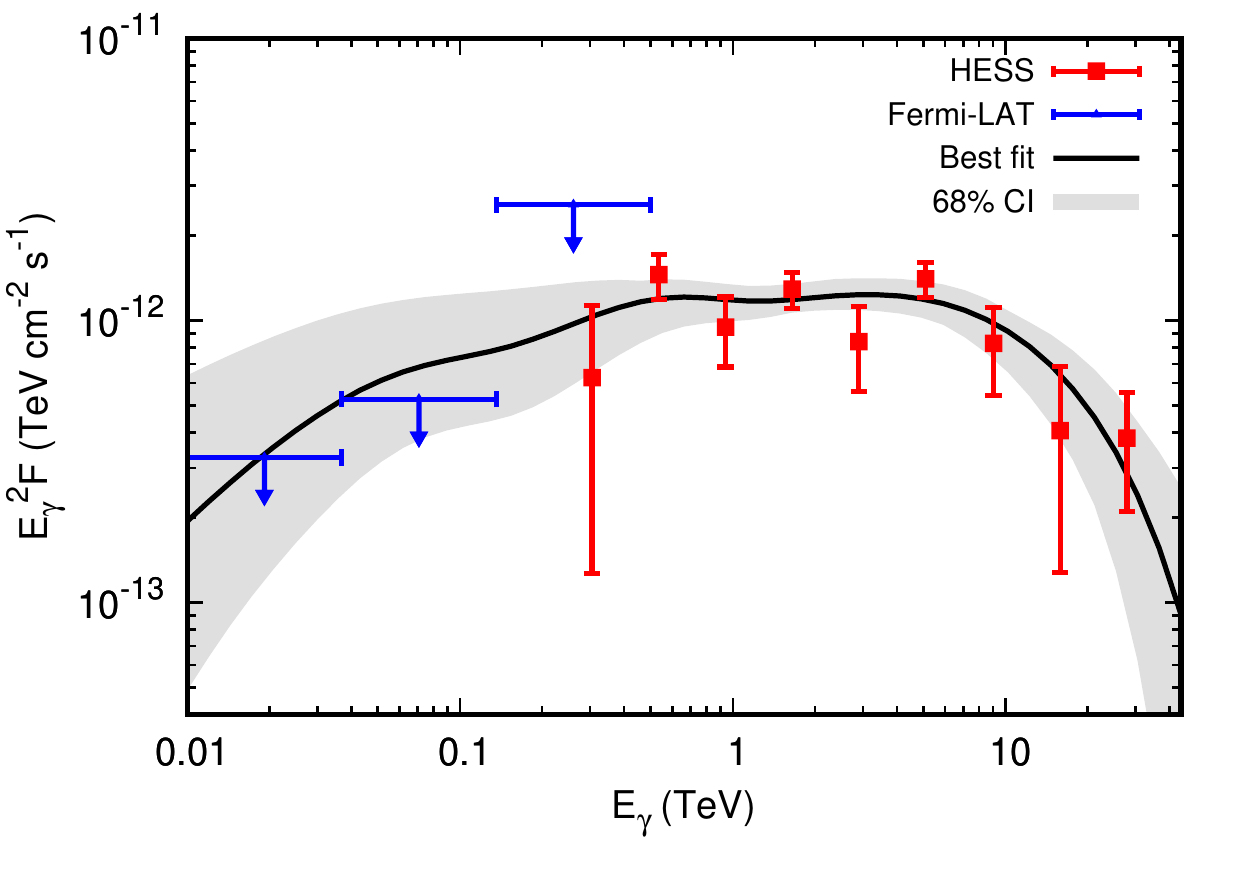}
\caption{The best-fit gamma-ray spectrum to the H.E.S.S. data \cite{Sheidaei:2011vg} using the basic model, compared with the H.E.S.S. and \textit{Fermi}-LAT data. The 68\% confidence interval of the fitted spectrum is also shown. Note the \textit{Fermi}-LAT ULs are not used in the fitting process.}
\label{fig:spec}
\end{figure}

The free parameters of the spectral fit are $p$, $E_c$, and $\eta$. The normalization of the injection spectrum is mainly determined by $\eta$. We compute $\int_{\ang{0}}^{\ang{0.3}}S(\theta,E_\gamma)2\pi\theta d\theta$ to fit the spectrum given by Ref.~\cite{Sheidaei:2011vg}. We apply the MULTINEST software \cite{Feroz:2013hea} for the data fit. The fitting result is presented in Fig.~\ref{fig:spec}. The best-fit values and posterior confidence intervals of the parameters are listed in Table~\ref{tab:para}. 

The best-fit conversion efficiency is 6.6\%. For the Geminga halo, the conversion efficiency is $\approx5\%$ to interpret the HAWC data if the injection spectrum also takes the form of Eq.~(\ref{eq:inj}) \cite{2021arXiv210707395B}, which is very similar to the needed value for HESS~J1831$-$098. The cutoff energy $E_c$ is constrained to around 50~TeV, while we may not rule out the possibility of a broken power-law injection spectrum with the H.E.S.S. data. Experiments working in higher energy regions like the 1.3~km$^2$ array of the large high-altitude air shower observatory (LHAASO-KM2A) \cite{Ma:2022aau} may provide a clearer judgment on the high-energy spectral features of HESS~J1831$-$098. The best-fit power-law index of 0.88 indicates a very hard injection spectrum, similar to that derived from the observations of the Geminga x-ray PWN \cite{Posselt:2016lot}, but significantly harder than the typical electron spectrum of PWNe, i.e., $p=1.5$ \cite{2017SSRv..207..175R}. The CR positron spectrum may also constrain the injection spectrum of PWNe. Ref.~\cite{Bitter:2022uqj} argues that the typical injection spectral index should be $p=1.5-1.7$ to interpret the AMS-02 data.

As can be seen in Fig.~\ref{fig:spec}, the flux ULs of \textit{Fermi}-LAT also require a very hard injection spectrum under the basic model. The \textit{Fermi}-LAT ULs are obtained using the disk template, which could be conservative constraints. The UL around 100~GeV is slightly lower than the best-fit spectrum to the H.E.S.S. data, indicating the injection spectral index may need to be even smaller than $0.88$. This is very similar to the case of LHAASO~J0621$+$3755: a more sophisticated propagation model is required to consistently interpret the LHAASO-KM2A and \textit{Fermi}-LAT data, otherwise an unreasonably hard injection spectrum will be required \cite{Fang:2021qon}. 

\section{Discussion}
\label{sec:discuss}
\begin{figure}[t]
\centering
\includegraphics[width=0.72\textwidth]{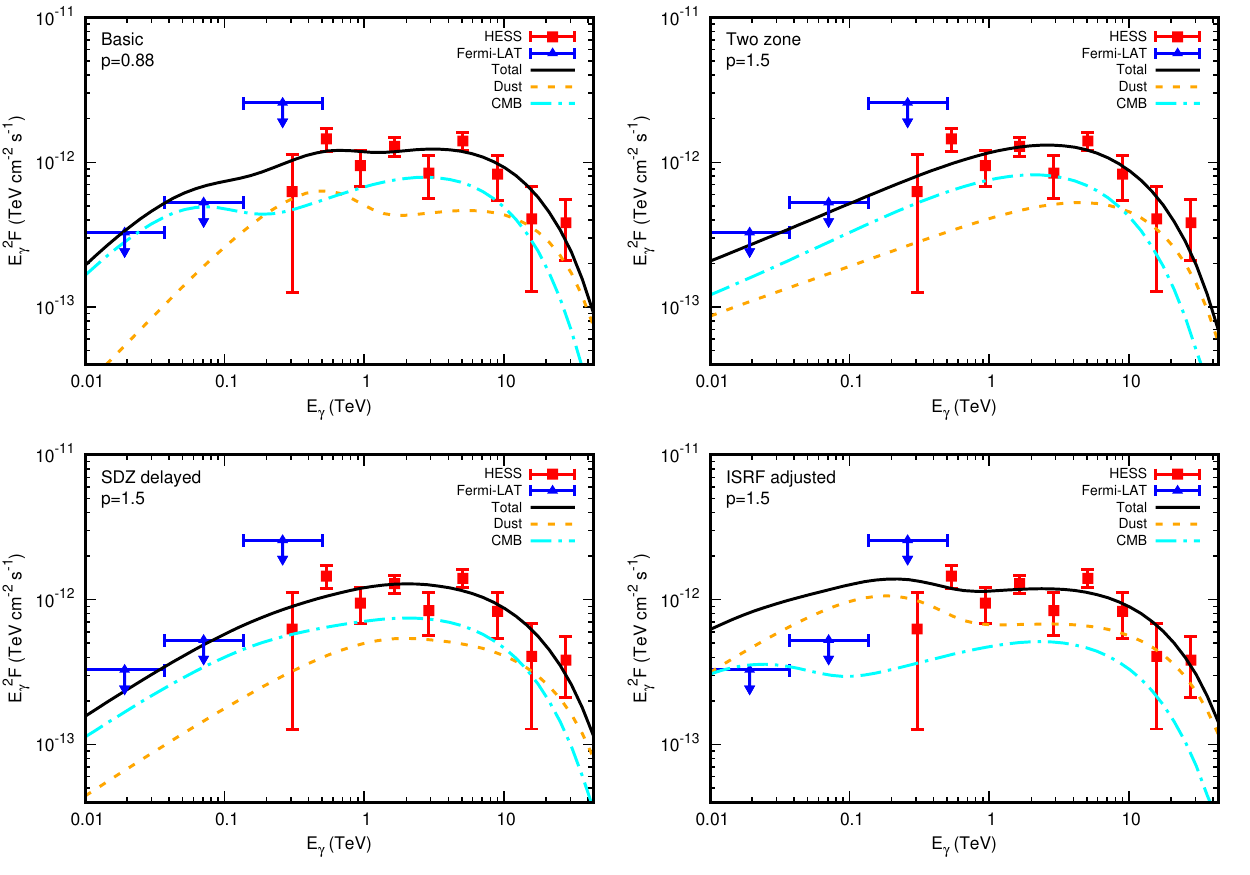}
\caption{The top-left graph is the same as Fig.\ref{fig:spec}, while the two sub-components of the gamma-ray spectrum are also shown. The other three graphs show models different from the basic model, which try to interpret the observations under a typical injection spectral index of $p=1.5$. Top right: the two-zone diffusion model, where the slow-diffusion zone has a finite size of 40~pc. Bottom left: a time delay of 50~kyr is assumed for the formation of the slow-diffusion zone (SDZ). Bottom right: the energy density of the dust photons is 2 times larger than the basic model.}
\label{fig:4spec}
\end{figure}

In this section, we discuss the possible scenarios that may consistently interpret the H.E.S.S. and \textit{Fermi}-LAT observations under a typical power-law index of the electron injection spectrum. The IC gamma-ray spectrum mainly consists of two components generated by scattering the dust and CMB photons, respectively. The top left of Fig.~\ref{fig:4spec} separately shows the dust and CMB components of the best-fit spectrum in Fig.~\ref{fig:spec}. Each component has a two-peak spectral feature, similar to the predicted CR positron spectrum at Earth generated by nearby middle-aged pulsars \cite{Fang:2019ayz}. The low-energy peak is dominantly produced by electrons injected in the early ages of the pulsar, while the high-energy peak is produced by the lately injected ones. Thus, if the number of the early injected electrons is reduced, the low-energy gamma-ray fluxes may meet the constraints of \textit{Fermi}-LAT, while the high-energy fluxes can still interpret the H.E.S.S. data.

The diffusion coefficient in the basic model is spatially and temporally independent, which is a commonly used but oversimplified assumption. The measured diffusion coefficient around the pulsar is two orders of magnitude smaller than the value inferred from the CR boron-to-carbon (B/C) ratio \cite{Aguilar:2016vqr}, which means that the slow diffusion cannot be representative in the Galaxy. As mentioned in Sec.~\ref{sec:model}, the slow-diffusion environment around the pulsar may either be self-excited or left by the parent SNR. Both the scenarios predict a size of $\sim50$~pc for the slow-diffusion zone \cite{Evoli:2018aza,Kun:2019sks,Mukhopadhyay:2021dyh}. The propagation outside this zone is assumed to be the fast diffusion indicated by the B/C ratio, which is known as the two-zone diffusion model \cite{Fang:2018qco}. The diffusion coefficient takes the form of 
\begin{equation}
 D(E_e, \mathbi{r})=\left\{
 \begin{aligned}
  & D_{\rm slow}(E_e)\,,\quad & |\mathbi{r}-\mathbi{r}_p|<r_* \\
  & D_{\rm fast}(E_e)\,,\quad & |\mathbi{r}-\mathbi{r}_p|\geq r_* \\
 \end{aligned}
 \right.\,,
 \label{eq:2zone}
\end{equation}
where $r_*$ is the size of the slow-diffusion zone, $D_{\rm fast}$ is taken from Ref.~\cite{Yuan:2017ozr}, and we adopt the best-fit diffusion coefficient obtained in Sec.~\ref{sec:hess} for $D_{\rm slow}$. Under the slow-diffusion condition, early injected electrons with $E_e=1$~TeV can propagate $\approx60$~pc away from the pulsar, larger than $r_*$. It means those electrons have already escaped from the slow-diffusion zone, and the low-energy gamma-ray fluxes should be significantly reduced compared with the basic model. Meanwhile, electrons with $E_e=100$~TeV can only propagate $\approx30$~pc owing to the short lifetime, which means that the high-energy gamma-ray spectrum is hardly affected by the two-zone diffusion assumption. The top right of Fig.~\ref{fig:4spec} presents a spectrum calculated with the two-zone diffusion model, where $p=1.5$ and $r_*=40$~pc. The low-energy spectral peaks of the two sub-components are smoothed out compared with the basic model, and the observations are well interpreted by the model. 

On the other hand, the slow-diffusion environment may not exist at the birth of the pulsar, which is suggested by both the self-generated and SNR-generated scenarios as discussed in Ref.~\cite{Fang:2022mdg}. If there is a time delay in the formation of the slow-diffusion zone, where the diffusion coefficient takes the form of 
\begin{equation}
 D(E_e, t)=\left\{
 \begin{aligned}
  & D_{\rm fast}(E_e)\,,\quad & t<t_* \\
  & D_{\rm slow}(E_e)\,,\quad & t\geq t_* \\
 \end{aligned}
 \right.\,,
 \label{eq:delay}
\end{equation}
early injected electrons will not be trapped around the pulsar, and the low-energy gamma-ray fluxes should be significantly suppressed. We assume $p=1.5$ and $t_*=50$~kyr to test this scenario and present the gamma-ray spectrum in the bottom left of Fig.~\ref{fig:4spec}. The result is similar to the two-zone diffusion case. A more realistic scenario could be the combination of these two models.

Apart from the propagation-revised models, a different assumption of the electron injection may also help to interpret the observations. The injection time profile is commonly assumed to share the same time dependency as the pulsar spin-down luminosity. However, the magnetic field of young PWNe could be much stronger than that of today, which means electrons may significantly lose their energies before escaping from PWNe. As a result, the injection power at early ages may be much weaker than the basic model, and the low-energy gamma-ray fluxes may be suppressed to interpret the \textit{Fermi}-LAT ULs. 

The top left of Fig.~\ref{fig:4spec} shows that the gamma-ray spectrum below $\approx100$~TeV is dominated by the CMB component. The straightforward idea may be that the conflict could be resolved if the proportion of the CMB component is suppressed, equivalent to a higher energy density of the dust photons than estimated in Sec.~\ref{sec:model}. However, the CMB component also has a significant contribution in $\approx1-10$~TeV. If the proportion of the CMB component is suppressed, the dust component is required to dominate the high-energy spectrum, which will also lead to an overproduction of low-energy gamma rays, as shown in the bottom right of Fig.~\ref{fig:4spec}.

\section{Conclusion and prospect}
\label{sec:conclu}
We study the likely pulsar halo HESS~J1831$-$098 with the H.E.S.S. and \textit{Fermi}-LAT observations. The HGPS provides morphology information to constrain the diffusion coefficient inside the pulsar halo, while the spectrum measurement of H.E.S.S. and \textit{Fermi}-LAT constrain the electron injection parameters. For the first time, we provide detailed evidence to prove that HESS~J1831$-$098 is a pulsar halo. HESS~J1831$-$098 meets all the criteria for a pulsar halo, including the position coincidence with a powerful pulsar, an appropriate pulsar age, a reasonable energy requirement, and a large size ratio between the gamma-ray halo and x-ray PWN. The derived diffusion coefficient is $9.0\times10^{27}$~cm$^2$~s$^{-1}$, comparable to the other known pulsar halos. The needed conversion efficiency from the pulsar spin-down energy to the electron injection energy is $6.6\%$, similar to the Geminga halo under the same assumption.

A single electron population with an exponentially-cutoff power-law injection spectrum can interpret the H.E.S.S. spectrum well. It is different from the case of the Geminga halo, which may be due to the difference in the injection process. The cutoff energy is constrained to around 50~TeV, which will be crosschecked by LHAASO and HAWC. A power-law term with $p\lesssim1$, much harder than the typical injection spectrum, is needed to interpret the H.E.S.S. spectrum and \textit{Fermi}-LAT ULs under the basic model. This is similar to the case of another pulsar halo, LHAASO~J0621$+$3755. The diffusion coefficient is spatially and temporally independent in the basic model, which could be oversimplified. Considering the possible mechanisms of the slow-diffusion zone, we adopt the two-zone diffusion model and the time-delayed slow-diffusion model. Under a typical power-law index of $p=1.5$, both of them can suppress the low-energy fluxes under the \textit{Fermi}-LAT ULs while interpreting the high-energy H.E.S.S. data. Assuming a time profile of electron injection different from the pulsar spin-down luminosity may also help solve this problem.

The morphology measurement given by HGPS is the integrated fluxes above 1~TeV. An energy-dependent measurement for the pulsar halo is essential to test sophisticated propagation models. For example, low-energy electrons are more likely to escape from the slow-diffusion region in the two-zone diffusion model. As a result, the gamma-ray extension in the low-energy range will be determined by the size of the slow-diffusion zone, which means the energy dependence of the extension may deviate from that predicted by the one-zone diffusion model. The energy-dependent morphology measurement is also essential to disentangle injection parameters from propagation parameters.

The diffusion coefficient measured by HESS~J1831$-$098 is comparable to other known pulsar halos. However, the spin-down luminosity of PSR~J1831$-$0952 is dozens of times larger than the other associated pulsars. The independence of the diffusion coefficient and the injection power indicates that the growth of the magnetic fluctuations may have reached saturation for the self-excited scenario of the slow-diffusion environment; otherwise, the slow diffusion may originate from external sources. The trend needs to be confirmed by a larger sample of pulsar halos.

\begin{acknowledgments}
This work is supported by the National Natural Science Foundation of China under Grants No. 12175248, No. 12105292, and No. U1738209.
\end{acknowledgments}

\bibliography{references}

\end{document}